\documentclass[12pt]{article}
\usepackage{amsmath,amssymb,latexsym,epsf,epsfig,graphicx,tikz}

\begin{document}

\newcommand{\sect}[1]{\setcounter{equation}{0}\section{#1}}
\renewcommand{\theequation}{\thesection.\arabic{equation}}
\newcommand{\prt}{\partial}
\newcommand{\II}{\mbox{${\mathbb I}$}}
\newcommand{\CC}{\mbox{${\mathbb C}$}}
\newcommand{\RR}{\mbox{${\mathbb R}$}}
\newcommand{\QQ}{\mbox{${\mathbb Q}$}}
\newcommand{\ZZ}{\mbox{${\mathbb Z}$}}
\newcommand{\NN}{\mbox{${\mathbb N}$}}
\def\G{\mathbb G}
\def\UU{\mathbb U}
\def\S{\mathbb S}
\def\tS{\widetilde{\mathbb S}}
\def\V{\mathbb V}
\def\tV{\widetilde{\mathbb V}}
\newcommand{\D}{{\cal D}}
\def\hint{H_{\rm int}}
\def\R{{\cal R}}

\newcommand{\rd}{{\rm d}}
\newcommand{\diag}{{\rm diag}}
\newcommand{\U}{{\cal U}}
\newcommand{\K}{{\mathcal K}}
\newcommand{\cP}{{\cal P}}
\newcommand{\dQ}{{\dot Q}}
\newcommand{\dS}{{\dot S}}

\newcommand{\ph}{\varphi}
\newcommand{\phd}{\widetilde{\varphi}} 
\newcommand{\phs}{\varphi^{(s)}}
\newcommand{\phb}{\varphi^{(b)}}
\newcommand{\phds}{\widetilde{\varphi}^{(s)}}
\newcommand{\phdb}{\widetilde{\varphi}^{(b)}}
\newcommand{\lambdad}{\widetilde{\lambda}}
\newcommand{\tx}{\widetilde{x}} 
\newcommand{\etat}{\widetilde{\eta}}
\newcommand{\phl}{\varphi_{i,L}}
\newcommand{\phr}{\varphi_{i,R}}
\newcommand{\phz}{\varphi_{i,Z}}
\newcommand{\mur}{\mu_{{}_R}}
\newcommand{\mul}{\mu_{{}_L}}
\newcommand{\muv}{\mu_{{}_V}}
\newcommand{\mua}{\mu_{{}_A}}
\newcommand{\mut}{\widetilde{\mu}}

\def\a{\alpha}
 
\def\A{\mathcal A} 
\def\H{\mathcal H} 
\def\U{\mathcal U} 
\def\E{\mathcal E} 
\def\C{\mathcal C} 
\def\L{\mathcal L} 
\def\O{\mathcal O}
\def\I{\mathcal I}
\def\der{\partial }
\def\mis{{\frac{\rd k}{2\pi} }}
\def\ri{{\rm i}}
\def\xt{{\widetilde x}}
\def\ft{{\widetilde f}}
\def\gt{{\widetilde g}}
\def\qt{{\widetilde q}}
\def\tt{{\widetilde t}}
\def\tmu{{\widetilde \mu}}
\def\prt{{\partial}}
\def\tr{{\rm Tr}}
\def\inc{{\rm in}}
\def\out{{\rm out}}
\def\li{{\rm Li}}
\def\e{{\rm e}}
\def\eps{\varepsilon}
\def\k{\kappa}
\def\v{{\bf v}}
\def\ebf{{\bf e}}
\def\abf{{\bf A}}
\def\fa{{\mathfrak a}} 


\newcommand{\finprf}{\null \hfill {\rule{5pt}{5pt}}\\[2.1ex]\indent}

\pagestyle{empty}
\rightline{November 2014}

\begin{center}
{\Large\bf Energy transmutation in nonequilibrium\\ quantum systems}
\\[2.1em]

\bigskip

{\large Mihail Mintchev}\\ 
\medskip 
{\it  
Istituto Nazionale di Fisica Nucleare and Dipartimento di Fisica, Universit\`a di
Pisa, Largo Pontecorvo 3, 56127 Pisa, Italy}
\bigskip 

{\large Luca Santoni}\\ 
\medskip 
{\it  
Scuola Normale Superiore, Piazza dei Cavalieri 7, 56126 Pisa, Italy}
\bigskip 

{\large Paul Sorba}\\ 
\medskip 
{\it  
LAPTh, Laboratoire d'Annecy-le-Vieux de Physique Th\'eorique, 
CNRS, Universit\'e de Savoie,   
BP 110, 74941 Annecy-le-Vieux Cedex, France}
\bigskip 

\end{center}
\begin{abstract} 
We investigate the particle and heat transport in quantum junctions with the geometry of star graphs. 
The system is in a nonequilibrium steady state, characterized by the different temperatures and 
chemical potentials of the heat reservoirs connected to the edges of the graph. We explore 
the Landauer-B\"uttiker state and its orbit under parity and time reversal transformations. Both particle number 
and total energy are conserved in these states. However, the heat and chemical potential energy are in general 
not separately conserved, which gives origin to a basic process of energy transmutation among them. 
We study both directions of this process in detail, introducing appropriate efficiency coefficients. 
For scale invariant interactions in the junction our results are exact and explicit. They cover the whole 
parameter space and take into account all nonlinear effects. The energy transmutation depends on 
the particle statistics.

\end{abstract}
\bigskip 
\medskip 
\bigskip 

\vfill
\rightline{LAPTH-107/14}
\rightline{IFUP-TH 9/2014}
\newpage
\pagestyle{plain}
\setcounter{page}{1}

\section{Introduction} 
\medskip 

The study of non-equilibrium quantum systems is among the most rapidly expanding areas of theoretical physics. 
Triggered by the remarkable experimental progress in manipulating trapped ultra-cold atomic gases, there is recently 
great interest in the search for universal properties of such systems (see e.g. \cite{pssv} for a review). 
Much attention is devoted to the behavior of quantum systems after a quench and in particular, 
on the nature of the equilibrium state which is approached in this case. Other interesting studies concern 
the impact of both internal and space-time symmetries (e.g. scale invariance) on 
the quantum transport in Non-Equilibrium Steady States (NESS). 

In this paper we investigate some general features of the energy transport in NESS's, representing 
non-equilibrium extensions of a Gibbs State (GS). The physical models we focus on are 
schematically shown in Fig. 1. We are dealing with a multicomponent system represented by the 
$n\geq 2$ semi-infinite leads (edges) $L_i$ of a star graph $\Gamma$. Each lead $L_i$ is 
attached at infinity to a heat reservoir $R_i$. The interaction between the 
$n$ components of the system is localized in the vertex of the graph and is defined by a scattering matrix $\S$. 
Each heat reservoir $R_i$ is described by a GS characterized by inverse temperature $\beta_i$ and chemical potential 
$\mu_i$. From Fig. 1 it is evident that the system is away from equilibrium if $\S$ admits a 
nontrivial transmission coefficient between two reservoirs $R_i$ and $R_j$ with 
$(\beta_i,\mu_i)\not=(\beta_j,\mu_j)$.

\begin{figure}[h]
\begin{center}
\begin{picture}(550,110)(20,255) 
\includegraphics[scale=0.8]{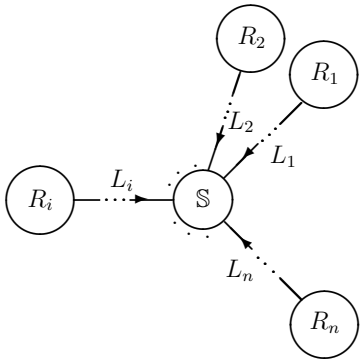}
\end{picture} 
\end{center}
\caption{A star graph $\Gamma$ with scattering matrix $\S$ at the vertex 
and leads $L_i$ connected at infinity to thermal reservoirs $R_i$.} 
\label{junction}
\end{figure}

We start our investigation by shortly reviewing the explicit construction of a NESS $\Omega_{\beta, \mu}$, 
induced by the scattering matrix $\S$ from the tensor product of the GS's relative to the heat reservoirs $R_i$. 
The state $\Omega_{\beta, \mu}$ is fully determined by $\S$ and $(\beta_i,\mu_i)$. 
We show that $\Omega_{\beta, \mu}$ is not invariant under parity and time-reversal 
transformations and generates therefore a nontrivial orbit $O_{\beta,\mu}$, consisting 
of four nonequivalent NESS's. The total energy of the system in these 
states has two components: heat energy (parametrized by $\beta_i$) and chemical potential 
energy (parametrized by $\mu_i$). Provided that the 
dynamics of the system is invariant under time translations, 
the total energy is conserved. For generic values of $(\beta_i,\mu_i)$ 
the two energy components are however not 
separately conserved and an energy transmutation occurs. The process is controlled by a single parameter $\dQ$, 
describing the heat flow in the junction. For $\dQ<0$ heat energy is transformed in chemical potential energy. 
The opposite transformation takes place for $\dQ>0$. 

The regime $\dQ<0$ has been investigated mostly by means of linear response 
theory. For the details we refer to the review paper \cite{bcps} and the references therein, 
observing only that in these studies the junction 
is usually compared to a heat engine, whose efficiency represents the main point 
of interest. This comparison is very suggestive, keeping always in 
mind that the junction acts actually as a converter between 
heat and chemical potential energy. The case $\dQ>0$ is quite subtle and has been explored only partially 
in the domain where the junction can be compared to a refrigerator or heat pump. 
The main goal of the present paper is to study the system in Fig. 1 as an energy converter, thus providing a 
systematic and unified description of both regimes $\dQ\lessgtr 0$, which are characterized by 
appropriate efficiency coefficients. Since the energy transport in the junction 
is strongly influenced by nonlinear phenomena, the well known linear response approximation 
is not suitable for our investigation because it covers only a little part of the parameter space. For this reason 
a fundamental objective of our study is to simplify as much as possible the dynamics in order to solve the 
problem in exact form, preserving at the same time a nontrivial quantum transport. 
In this respect we demonstrate that assuming free propagation along $L_i$ and scale-invariant 
point-like interactions in the vertex is enough for this purpose. 
We study both fermionic and bosonic systems, showing that the particle statistics affects 
the energy transmutation. The detailed comparison between the two cases is very instructive, 
because the bosonic case is poorly investigated. The impact of parity and time reversal 
on the quantum transport is discussed in detail as well. 

We proceed as follows. In the next section we briefly recall the construction of the NESS $\Omega_{\beta, \mu}$ 
and its orbit $O_{\beta, \mu}$ under parity and time reversal transformations. 
In section 3 we fix the Schr\"odinger dynamics on the leads and the interaction at the vertex of $\Gamma$. 
Studying the local conserved currents and the relative densities, we show that the junction converts 
heat energy in chemical potential energy or vice versa, depending on the values of the 
parameters characterizing the reservoirs $R_i$. 
The basic features of this process of energy transmutation 
are investigated in section 4, where we introduce the efficiency 
coefficients for both cases $\dQ\lessgtr 0$. We describe here also the scale invariant (critical) interactions 
in the junction. The efficiency of the quantum transport for a system with two leads is investigated 
in section 5 both for the LB state $\Omega_{\beta, \mu}$ and the orbit $O_{\beta, \mu}$. We also  
relate here our approach to some exact results, concerning other systems and NESS's. In section 6 we describe the 
main features of a junction with 3 leads. Section 7 is devoted to the bosonic case. The conclusions 
and some future developments form the content of Section 8. The appendices collect some technical 
results about the Onsager matrix and the 3-lead junction.

\bigskip

\section{Non equilibrium states on $\Gamma$} 
\medskip 

Following the pioneering work of Landauer \cite{la-57} and B\"uttiker \cite{bu-86}, non-equilibrium 
systems of the type in Fig. 1 have been extensively investigated. 
In this section we briefly recall an explicit quantum field theory construction \cite{Mintchev:2011mx} 
of the Landauer-B\"uttiker (LB) state $\Omega_{\beta, \mu}$, which adapts to the star graph case 
some general ideas \cite{els-96}-\cite{st-06} about NESS's. This construction is very convenient 
for deriving correlation functions and for studying the orbit $O_{\beta, \mu}$ of $\Omega_{\beta, \mu}$ 
under parity and time reversal transformations. As mentioned in the introduction, 
the elements of this orbit represent new nonequivalent NESS's with interesting physical properties. 

The first step in constructing the LB state $\Omega_{\beta, \mu}$ is to describe the asymptotic dynamics 
at $t=-\infty$ (i.e. before the interaction) in terms of the operators 
\begin{equation}
\{a^*_i(k),\, a_i(k)\, :\, k >0, \, i=1,...,n\}\, , 
\label{in}
\end{equation}
which create and annihilate the particle excitations with momentum $k$ in the reservoir $R_i$. 
In agreement with the orientation of the leads in Fig.1, the condition $k>0$ implies that (\ref{in}) create and annihilate 
incoming  particles. For fermionic systems (\ref{in}) generate therefore 
an incoming Canonical Anti-commutation Relations (CAR) algebra $\A^\inc$. We denote by $\Omega_{\beta_i,\mu_i}$ 
be the GS associated with the heat reservoir $R_i$ (see e.g. \cite{BR}) and consider the tensor product 
$\otimes_{i=1}^n \Omega_{\beta_i,\mu_i}$. 

The next step is to relate $\A^\inc$ with the 
CAR algebra of outgoing excitations $\A^\out$, 
generated still by the creation and annihilation operators (\ref{in}), 
but with $k<0$. For this purpose we have to 
introduce the interaction, connecting different heat reservoirs and driving the system away from  
equilibrium. We consider the simple case where the incoming 
particles propagate freely along the leads towards the vertex of the graph, where they are 
reflected or transmitted with some probability in the rest of the graph. This process is 
codified in the reflection-transmission equations  
\begin{equation} 
a_i(k) = \sum_{j=1}^n \S_{ij} (k) a_j (-k) \, , \qquad 
a^\ast_i (k) = \sum_{j=1}^n a^\ast_ j(-k) \S^*_{ji} (k) \qquad k<0 \, ,    
\label{constr1}
\end{equation} 
which relate $\A^\inc$ and $\A^\out$. Here $\S(k)$ is the scattering matrix 
describing the point-like interaction in the vertex of the graph. 
We assume unitarity and Hermitian analyticity 
\begin{equation}
\S(k) \S(k)^* = \II \, , \qquad \S(k)^*=\S(-k) \, , 
\label{unitha}
\end{equation} 
the star ${}^*$ indicating Hermitian conjugation. These conditions imply 
that $\S(k) \S(-k) = \II$, which ensures the consistency of the constraints (\ref{constr1}). 

The whole algebra $\A$, generated by polynomials involving generators of both $\A^\inc$ and $\A^\out$, is 
a deformed\footnote{CAR algebra 
deformations of the type (\ref{rta2}) have been studied previously in the context of 
one-dimensional integrable systems with boundaries \cite{Liguori:1996xr} or 
defects \cite{Mintchev:2003ue}.} CAR algebra, 
where $[a_i(k)\, ,\, a_j(p)]_+ = [a^*_i (k)\, ,\, a^*_j (p)]_+ = 0$ and 
\begin{equation}
[a_i(k)\, ,\, a^*_j (p)]_+ = 2\pi [\delta (k-p)\delta_{ij} + \S_{ij}(k)\delta(k+p)] \, . 
\label{rta2}
\end{equation}  
Because of (\ref{unitha}), the right hand side of (\ref{rta2}) defines 
the kernel of an integral projection (instead of the usual identity) operator.
Hermitian analyticity implies \cite{Liguori:1996xr} that 
the $*$-operation in $\A$ is a conjugation.  

At this point the LB state $\Omega_{\beta, \mu}$ is 
the extension of $\otimes_{i=1}^n \Omega_{\beta_i,\mu_i}$ from 
$\A^\inc$ to the whole algebra $\A$, performed by linearity via  
the reflection-transmission relations (\ref{constr1}). This construction, which may look at the 
first sight a bit abstract, is in fact very efficient for deriving the correlation functions 
defining the LB representation $\H_{\rm LB}$ of $\A$. We stress that $\H_{\rm LB}$ describes non-equilibrium physics 
and is not equivalent to the Fock and Gibbs representations $\H_{\rm F}$ and $\H_{\rm G}$ of $\A$, 
known \cite{Liguori:1996xr,Mintchev:2004jy} from the equilibrium case. Denoting by $(\cdot\, ,\, \cdot)$ the scalar 
product in the Hilbert space $\H_{\rm LB}$, one has \cite{Mintchev:2011mx} 
\begin{eqnarray} 
(\Omega_{\beta, \mu}\, ,\, a_j^*(p)a_i(k) \Omega_{\beta, \mu}) \equiv 
\langle a_j^*(p)a_i(k)\rangle_{\beta, \mu} = 
\qquad \qquad \qquad \nonumber \\ 
2\pi \delta (k-p)\left [\theta(k)\delta_{ij}d_i(k) + 
\theta(-k)\sum_{l=1}^n \S_{il}(k)\, d_l(-k)\, \S^*_{lj}(k)\right ]  \qquad 
\nonumber \\
+2\pi \delta (k+p)\left [\theta(k) d_i(k) \S^*_{ij}(-k) + \theta(-k)\S_{ij}(k) d_j(-k) \right ] \, ,
\qquad \; \; \,  
\label{cor1}
\end{eqnarray} 
were 
\begin{equation} 
d_i(k) = \frac{\e^{-\beta_i \left [\omega_i (k) -\mu_i\right ]}}
{1+ \e^{-\beta_i \left [\omega_i (k) -\mu_i \right ]}} 
\label{fbd1} 
\end{equation} 
is the Fermi distribution in the heat reservoir $R_i$ with dispersion relation $\omega_i(k)$. 
Notice that the construction allows for different dispersion relations in the different reservoirs.  

The explicit form of $\langle a_i(k)a_j^*(p)\rangle_{\beta, \mu}$ is 
obtained from (\ref{cor1}) by the substitution 
\begin{equation} 
d_i(k) \longmapsto c_i(k) =  
 \frac{1}{1+ \e^{-\beta_i \left [\omega_i (k) -\mu_i\right ]}} \, . 
\label{fbd2} 
\end{equation} 
As well known, employing the CAR algebra, one can express  
a generic $n$-point correlation function as a polynomial of the two-point 
correlator (\ref{cor1}). The non-equilibrium features of the correlation functions are encoded 
in the mixing of the Fermi distributions, associated with the different heat reservoirs $R_i$, via the scattering 
matrix $\S$, which is evident already from (\ref{cor1}). 

An advantage of the above framework is that it allows to investigate directly the 
behavior of the LB state $\Omega_{\beta, \mu}$ under 
parity and time reversal. We first recall that these operations are 
implemented by a unitary operator $P$ and an anti-unitary operator $T$, 
acting in the algebra $\A$ in the standard way: 
\begin{equation}
P a_i(k) P^{-1} = \chi_P\,  a_i(-k)\, ,\qquad Ta^*_i(k) T^{-1} = \chi_T\,  a_i(-k)\, , 
\qquad |\chi_P|=|\chi_T|=1\, . 
\label{pt}
\end{equation}
Apart from the multiplicative phase factors, the action of $P$ and $T$ seems to be the same, 
but one should remember that $P$ is a linear operator, whereas $T$ is anti-linear. 
Using the explicit form (\ref{cor1}) of the two-point function and (\ref{pt}), one can explicitly verify 
that 
\begin{equation}
\left (X\Omega_{\beta , \mu}\, ,\,  a_j^*(p) a_i(k) X\Omega_{\beta , \mu} \right ) \equiv 
\langle a_j^*(p) a_i(k)  \rangle^X_{\beta, \mu} \not= \langle a_j^*(p)a_i(k)\rangle_{\beta, \mu}\, , 
\qquad X= P,\, T,\, PT\, ,
\label{xcorr}
\end{equation} 
if $(\beta_i,\mu_i)\not=(\beta_j,\mu_j)$ for some reservoirs. Therefore, 
$\Omega_{\beta, \mu}$ is not invariant under parity and time reversal and 
generates the nontrivial orbit 
\begin{equation}
O_{\beta,\mu}=\{\Omega_{\beta, \mu},\, \Omega^X_{\beta, \mu} = X \Omega_{\beta, \mu}\, :\, X= P,\, T,\, PT\}\, . 
\label{orbit}
\end{equation} 
The property (\ref{xcorr}) captures the fundamental difference of the 
LB state $\Omega_{\beta, \mu}$ with respect to the Fock vacuum $\Omega_{\rm F}\in \H_{\rm F}$ and the 
Gibbs state $\Omega_{\rm G}\in \H_{\rm G}$, which are both $P$- and $T$-invariant. 
The result (\ref{xcorr}) suggests also that the four states in the orbit $O_{\beta,\mu}$ have 
different transport properties, which is confirmed by the analysis in section 5.2 below. 
Since time reversal symmetry is the quantum counterpart of classical reversibility, 
one can interpret the breakdown of this symmetry in the LB representation $\H_{\rm LB}$ 
as quantum irreversibility. 

We conclude by observing that above construction can be easily generalized to bosons by 
replacing the CAR algebra with a Canonical Commutation Algebra (CCA), which implies 
the substitution of the Fermi distribution in (\ref{cor1}) with the Bose distribution: 
\begin{equation} 
d_i(k) \longmapsto b_i(k) =  
\frac{\e^{-\beta_i \left [\omega_i (k) -\mu_i\right ]}}{1- \e^{-\beta_i \left [\omega_i (k) -\mu_i\right ]}}\, . 
\label{bd2} 
\end{equation} 
As one can expect on general grounds and as shown in section 7, the quantum transport is 
influenced by the statistics.

\bigskip 

\section{The Schr\"odinger junction} 
\medskip 

Quantum systems away from equilibrium behave usually in a complicated way. In most of the cases the 
linear response or other approximations are not enough for fully describing the complexity 
of this behavior. For this reason the existence of models which incorporate the main non-equilibrium 
features, while being sufficiently simple to be analyzed exactly, is conceptually very important. 
An interesting family of such models 
in $s$ space dimensions is characterized by requiring that the interaction, 
which drives the system away from equilibrium, 
is localized on a $(s-1)$-dimensional sub-manifold, whereas the propagation 
in the complementary orthogonal direction is free. 
In what follows we will consider a special case of this scenario, focussing on a one-dimensional space 
with the geometry of a star graph $\Gamma$. Each point $P\in \Gamma$ 
is parametrized by the coordinates $\{(x,i)\, ,:\, x<0,\, i=1,...,N\}$, where $|x|$ is the distance of $P$ from the vertex and 
$i$ labels the lead. Since $s=1$, the interaction is implemented by a point-like defect 
in the vertex of $\Gamma$ and the propagation along the leads $L_i$ is free. We assume in 
this paper that it is governed by the Schr\"odinger equation  
\begin{equation}
\left (\ri \prt_t +\frac{1}{2m} \prt_x^2\right )\psi (t,x,i) = 0\, , 
\label{eqm1}
\end{equation} 
but other types of time evolution can be considered 
\cite{Mintchev:2011mx,Mintchev:2012pe} in the same way. The field $\psi$ is complex and the system has 
a global $U(1)$-invariance, generating particle number conservation. 
In the fermionic case $\psi$ satisfies the standard equal-time CAR's. 

The scattering matrix $\S$ in the vertex is fixed by requiring that the bulk Hamiltonian $-\prt_x^2$ admits a 
self-adjoint extension on the whole graph. These extensions are defined \cite{ks-00}-\cite{k-08} 
by 
\begin{equation} 
\lim_{x\to 0^-}\sum_{j=1}^n \left [\lambda (\II-\UU)_{ij} +\ri (\II+\UU)_{ij}\prt_x \right ] \psi (t,x,j) = 0\, , 
\label{bc1} 
\end{equation} 
where $\UU$ is a $n\times n$ unitary matrix and $\lambda \in \RR$ is a 
parameter with dimension of mass. Eq. (\ref{bc1}) guaranties unitary time evolution of the 
system on the graph. The matrices $\UU=\II$ and $\UU=-\II$ define the Neumann and Dirichlet 
boundary conditions respectively. The explicit form of the scattering matrix 
is \cite{ks-00}-\cite{k-08} 
\begin{equation} 
\S(k) = 
-\frac{[\lambda (\II - \UU) - k(\II+\UU )]}{[\lambda (\II - \UU) + k(\II+\UU)]} \, .   
\label{S1}
\end{equation} 
The diagonal element $\S_{ii}(k)$ represents the reflection amplitude from the vertex on the lead $L_i$, whereas  
$\S_{ij}(k)$ with $i\not=j$ equals the transmission amplitude from $L_i$ to $L_j$. 
One easily verifies that (\ref{S1}) satisfies (\ref{unitha}) and therefore defines an algebra 
$\A$ of the type introduced in the previous section. Moreover, $\S(k)$ is a meromorphic function 
in the complex $k$-plane with finite number of simple poles on the imaginary axis. For 
simplicity we consider in this paper the case without bound states (poles in the upper half plane), 
referring for the general case to \cite{Mintchev:2004jy}, \cite{Bellazzini:2010gs}.  
In this case the solution of equation (\ref{eqm1}) is fixed 
uniquely by (\ref{bc1}) and takes the following simple form
\begin{equation}
\psi (t,x,i) = \sum_{j=1}^n \int_{0}^{\infty} 
\frac{dk}{2\pi }
\e^{-\ri \omega (k)t}\left [\e^{-\ri kx}\, \delta_{ij} + \e^{\ri kx}\, \S_{ij}(-k)\right ] a_j (k) \, , \qquad  \omega(k) = \frac {k^2}{2m} \, . 
\label{psi1} 
\end{equation}
As expected, the parity and time-reversal transformations (\ref{pt}) in the algebra $\A$ imply 
\begin{equation}
P \psi (t,x,i) P^{-1} = \chi_P \psi (t,-x,i)\, , \qquad  
T \psi (t,x,i) T^{-1} = \chi_T \psi (-t,x,i)\, . 
\label{pt1}
\end{equation} 

Let us describe now the basic local observables, whose behavior away from equilibrium is the main topic of this paper. 
The local particle density and relative current are 
\begin{equation}
j_t (t,x,i)=  \left [ \psi^* \psi \right ]  (t,x,i)\, ,   
\qquad 
j_x(t,x,i)= \frac{\ri }{2m} \left [ \psi^* (\partial_x\psi ) - 
(\partial_x\psi^*)\psi \right ]  (t,x,i) \, , 
\label{curr1}
\end{equation} 
The total energy density is 
\begin{equation}
\theta_{tt} (t,x,i) = -\frac{1}{4m} \left[ \psi^* \left (\partial_x^2 \psi \right )+
\left (\partial_x^2 \psi^* \right )\psi \right] (t,x,i) \, , 
\label{endens1} 
\end{equation}
with energy flow  
\begin{equation}
\theta_{xt} (t,x,i) = \frac{1}{4m} [\left (\partial_t \psi^* \right )\left (\partial_x \psi \right ) 
+ \left (\partial_x \psi^* \right )\left (\partial_t \psi \right ) \\ - 
\left (\partial_t \partial_x \psi^* \right ) \psi - 
\psi^*\left (\partial_t \partial_x \psi \right ) ](t,x,i) \, . 
\label{en1} 
\end{equation} 
The equations of motion lead to the local conservation laws 
\begin{equation} 
\left (\partial_t j_t - \partial_x j_x \right )(t,x,i) = 
\left (\partial_t \theta_{tt} - \partial_x \theta_{xt} \right )(t,x,i) = 0 \, . 
 \label{cons1} 
\end{equation} 
The relation between local conservation laws and the associated charges on a star graph 
has been investigated in \cite{Bellazzini:2008mn}. In the presence of a defect, like 
the junction in our case, the local conservation 
(\ref{cons1}) alone is not enough \cite{Bellazzini:2008mn} to ensure the 
conservation of the relative quantum numbers. One needs in 
addition the Kirchhoff rules 
\begin{equation} 
\sum_{i=1}^n j_x(t,0,i) = 0\, , \qquad 
\sum_{i=1}^n \theta_{xt}(t,0,i) = 0 \, .  
\label{KK1}
\end{equation} 
It is worth stressing that the explicit form (\ref{S1}) 
of the $\S$-matrix is fundamental for proving \cite{Mintchev:2011mx} the operator form (\ref{KK1}) 
of the Kirchhoff rules, which guaranties  
the charge conservation in the whole state space $\H_{\rm LB}$ of the system. 
Combining (\ref{cons1}) and (\ref{KK1}) one 
concludes that for these $\S$-matrices the particle number and the total energy 
in our system are conserved. 

The heat density $q_t$ in the lead $L_i$ is obtained (see e.g. \cite{call}) by subtracting 
from the total energy density the energy density  
relative to the chemical potential $\mu_i$, namely 
\begin{equation} 
q_t(t,x,i) = \theta_{tt} (t,x,i) - \mu_i j_t (t,x,i) \, . 
\label{hd}
\end{equation} 
Accordingly, the heat current is 
\begin{equation} 
q_x(t,x,i) = \theta_{xt} (t,x,i) - \mu_i  j_x (t,x,i) \, .  
\label{hc}
\end{equation} 
Local heat conservation  
\begin{equation} 
\left (\partial_t q_t - \partial_x q_x \right )(t,x,i) = 0 \, , 
\label{cons2} 
\end{equation}  
is a direct consequence of (\ref{cons1}), but in general the relative Kirchhoff rule is not at all 
satisfied. In fact, a key observation is that the heat current 
obeys the operator Kirchhoff rule if and only if $\mu_i = \mu_j$ for all $i,j=1,...,N$. 
Otherwise, the heat current violates the Kirchhoff rule and the 
heat energy is therefore not conserved. The chemical potential energy shares the same property, 
because the total energy is conserved. 

Summarizing, if $\mu_i \not= \mu_j$ for some $i$ and $j$, 
the system converts heat energy in chemical potential energy or vice versa. This is the basic 
physical process which takes place in the junction. In what follows we will study in detail this 
phenomenon of energy transmutation. 

\bigskip 

\section{Transport in the state $\Omega_{\beta,\mu}$}

\subsection{Currents and efficiency}

In order to study the non-equilibrium features of our system, we derive 
now the expectation values in the state $\Omega_{\beta,\mu}$ of the charge densities 
and currents introduced above.  Since these observables are quadratic, the basic input is 
the non-equilibrium two-point correlation function 
\begin{eqnarray}
\langle \psi^*(t_1,x_1,i) \psi (t_2,x_2,j)\rangle_{\beta, \mu} = 
\int_0^{\infty} \frac{\rd k}{2\pi} \e^{\ri \omega(k) t_{12}} 
\Bigl [\delta_{ji} d_i(k) \e^{\ri k x_{12}} + 
\quad 
\nonumber \\
d_j(k) \S_{ji}(k)  \e^{-\ri k \tx_{12}} + \S^\ast_{ji}(k) d_i(k) \e^{\ri k \tx_{12}} + 
\sum_{l=1}^n \S^\ast_{jl}(k) d_l(k) \S_{li}(k) \e^{-\ri k x_{12}} \Bigr ] , 
\label{corr11}
\end{eqnarray} 
following from (\ref{cor1},\ref{psi1}). Here $t_{12}=t_1-t_2$, $x_{12}=x_1-x_2$ and 
$\tx_{12}=x_1+x_2$. The invariance of (\ref{corr11}) under time translations 
explicitly confirms that the total energy 
of the system is conserved. Combining (\ref{corr11}) with (\ref{curr1},\ref{en1}) 
and performing the limits $t_1 \to t_2=t$ and $x_1 \to x_2 =x$ one gets the 
following current expectation values 
\begin{equation}
J_i^N \equiv \langle j_x(t,x,i) \rangle_{\beta, \mu} =
\sum_{j=1}^n \int_0^{\infty} \frac{\rd k}{2\pi} \frac{k}{m} \left [\delta_{ij} - |\S_{ij}(k)|^2 \right ] d_j(k)\, , 
\label{ccurr1}
\end{equation} 
\begin{equation}
J_i^E \equiv 
\langle \theta_{tx}(t,x,i) \rangle_{\beta, \mu} =
\sum_{j=1}^n \int_0^{\infty} \frac{\rd k}{2\pi} 
\frac{k}{m} \left [\delta_{ij} - |\S_{ij}(k)|^2 \right ] \omega(k) d_j (k)\, . 
\label{ecurr1}
\end{equation} 
The right hand sides of (\ref{ccurr1},\ref{ecurr1}) are the Landauer-B\"uttiker expressions \cite{la-57, bu-86, si-86} 
for the particle and energy currents in our case specified by the $\S$ matrix (\ref{S1}). This is the 
main reason for referring to $\Omega_{\beta,\mu}$ as the LB state. Notice that for $\beta_1=\cdots =\beta_n$ and 
$\mu_1=\cdots =\mu_n$ the system is in equilibrium and the currents (\ref{ccurr1}, \ref{ecurr1}) vanish 
due to the unitarity of $\S$.   

{}From (\ref{hc}) one gets the heat current 
\begin{equation}
J_i^Q  = J_i^E -\mu_i J_i^N\, . 
\label{hc1}
\end{equation} 
As already observed, in general the heat and chemical potential currents $J_i^Q$ and $\mu_i J_i^N$ 
do not satisfy separately the Kirchhoff rule. In fact, the heat flow $\dot Q$ from the junction 
is given by 
\begin{equation} 
{\dot Q} + \sum_{i=1}^n J_i^Q = 0\, . 
\label{hf0}
\end{equation} 
Taking into account the Kirchhoff rule (\ref{KK1}) for the energy current one gets 
\begin{equation}
{\dot Q} = -\sum_{i=1}^n J_i^Q =  \sum_{i=1}^n \mu_i J_i^N \, ,
\label{hf1}
\end{equation}
If ${\dot Q} <0$ the junction transforms heat energy in chemical potential energy. The efficiency of this process 
can be characterized as follows. Let us denote by $\K_\out$ the subset of heat currents leaving the reservoirs $R_i$. 
With our choice for the orientation of the leads (see Fig. 1), these currents are positive. With this convention 
the efficiency of the junction to transform heat energy in chemical potential energy is defined by 
\begin{equation}
\eta = \frac{\sum_{i=1}^n J_i^Q}{\sum_{i\in \K_\out} J_i^Q}=\frac{-\dQ}{\sum_{i\in \K_\out} J_i^Q}\, ,
\label{eta1}
\end{equation}
which is a direct extension of the cases $n=2,3$ \cite{bcps, sns-14} to a generic $n$.  
Combining (\ref{hf1}) with ${\dot Q} <0$ one concludes that $\sum_{i\in \K_\out} J_i^Q>0$, which implies that $\eta$ 
is well defined and satisfies $0<\eta\leq 1$.  

In the regime ${\dot Q} >0$ of converting chemical potential energy to heat energy the junction 
is usually compared (see e.g. \cite{bcps}) to a refrigerator or a heat pump, thus involving in the description 
the known performance coefficients of these devises. We show below that this approach 
works in parts of the parameter space, but can not be applied globally 
to the whole domain with $\dQ>0$. In order to avoid this problem, for $\dQ>0$ we propose and adopt below 
the efficiency  
\begin{equation}
\etat = \frac{\sum_{i=1}^n \mu_iJ_i^N}{\sum_{i\in \L_\out} \mu_iJ_i^N} =\frac{\dQ}{\sum_{i\in \L_\out} \mu_iJ_i^N}\, , 
\label{etat1}
\end{equation}
where now the sum in the denumerator runs over the subset $\L_\out$ of outgoing (positive) chemical 
potential energy currents $\mu_iJ_i^N$. The general form of (\ref{etat1}) is analogous to that of (\ref{eta1}), 
but refers to the chemical potential energy. By construction $0<\etat\leq 1$ in this case as well. 

For $\dQ=0$ there is no energy transmutation. 

We conclude this subsection by observing that the heat flow in the system generates the following 
entropy production \cite{call} 
\begin{equation}
\dS = -\sum_{i=1}^n \beta_i J_i^Q \, . 
\label{entr1}
\end{equation} 
Employing (\ref{ccurr1},\ref{ecurr1}), $\dS$ can be written in the form  
\begin{equation} 
\dS = 
\sum_{i,j=1}^n \int_0^{\infty} \frac{\rd k}{2\pi} \frac{k}{m} |\S_{ij}(k)|^2 
\left [\sigma_i(k) -\sigma_j(k)\right ]d_j(k)\, , 
\label{nentr1}
\end{equation} 
with  
\begin{equation}
\sigma_i(k) = \beta_i\left [\omega(k) -\mu_i\right ] \, . 
\label{nentr2}
\end{equation} 
Except in section 5.3, we will always assume in this paper that the bulk theory and the heat reservoirs 
have the same dispersion relation, namely 
\begin{equation}
\omega_i(k) = \omega(k) \, . 
\label{nentr3}
\end{equation} 
With this assumption 
\begin{equation}
d_j(k) = \frac{1}{\e^{\sigma_j(k)}+1} \, . 
\label{nentr4}
\end{equation} 
Now, using that (\ref{nentr4}) is a strictly decreasing function of $\sigma_j$, one can prove \cite{gn} that 
the integrand of (\ref{nentr1}) is nonnegative\footnote {The elegant argument of \cite{gn} 
is based on the inequality $F(x)-F(y) \leq (x-y) f(y)$, where $F$ 
is any primitive of a strictly decreasing function $f$.}, implying the second law of thermodynamics $\dS \geq 0$ 
in the LB state for all $(\beta_i,\mu_i)$ and scattering matrices (\ref{S1}). 
As explained in section 5.2 below, this argument does not apply 
to the other states of the orbit $O_{\beta,\mu}$ and the entropy production there behaves indeed differently. 

Summarizing, the process of energy transmutation in the junction is controlled by the two parameters $\dQ$ and 
$\dS$ and is characterized by the efficiency coefficients $\eta$ and $\etat$. We derive in what follows the explicit 
form of these physical quantities, assuming that the interaction at the junction is scale invariant. 

\bigskip 

\subsection{The scale invariant junction} 
\medskip 

The great advantage of considering the non-equilibrium Schr\"odinger junction defined in section 3 is that it provides 
both exact and explicit results with nontrivial transport. In order to keep this property also after the integration in $k$, 
we select in what follows the scale-invariant elements among the scattering matrices (\ref{S1}). 
They preserve the basic features of the system, while being simple enough to allow the 
explicit $k$-integration in (\ref{ccurr1},\ref{ecurr1}). 
The requirement of scale invariance implies that the interaction 
is $k$-independent and leads to \cite{Calabrese:2011ru} 
\begin{equation}
\S = \U\, \S_d \, \U^*\, ,  \qquad \U\in U(n)\, , \qquad \S_d = {\rm diag}(\pm 1,\pm 1,...,\pm 1)\, .  
\label{sinvS}
\end{equation} 
The family (\ref{sinvS}) is the orbit of the diagonal matrix 
$\S_d$ under the adjoint action of $U(n)$. 
We can always enumerate the leads $L_i$ in such a way that the first $i$ eigenvalues of $\S$ are $+1$ 
and the remaining $n-i$ are $-1$. For $i=n$ and $i=0$ one gets $\S=\II$ and $\S=-\II$, 
which imply vanishing transport because the leads $L_i$ are disconnected. One has a nontrivial 
transport for $0<i<n$. 

The scattering matrices of the type (\ref{sinvS}) are in general not symmetric. 
One can easily prove however that the relative amplitudes are symmetric, namely 
\begin{equation}
|\S_{ij}|^2 = |\S_{ji}|^2 \, . 
\label{symS}
\end{equation}
This property of the critical points (\ref{sinvS}) simplifies considerably the study of junctions with $n\geq 3$ leads. 

\begin{figure}[h]
\begin{center}
\begin{picture}(700,40)(80,355) 
\includegraphics[scale=1]{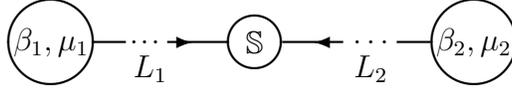}
\end{picture} 
\end{center}
\caption{Two heat reservoirs $(\beta_1,\mu_1)$ and $(\beta_2,\mu_2)$ connected by $\S$.} 
\label{junction2}
\end{figure}

The $k$-integration in (\ref{ccurr1},\ref{ecurr1}) with the constant scattering matrices (\ref{sinvS}) can be performed 
explicitly. One finds 
\begin{equation}
J^N_i  =
\frac{1}{2\pi} \sum_{j=1}^n \left [\delta_{ij} - |\S_{ij}|^2 \right ] 
 \frac{1}{\beta_j} \ln \left (1+\e^{\beta_j\mu_j} \right ) \, , 
\label{ccurr2}
\end{equation} 
\begin{equation}
J^E_i =
-\frac{1}{2\pi} \sum_{j=1}^n \left [\delta_{ij} - |\S_{ij}|^2 \right ] 
 \frac{1}{\beta_j^2} \li_2 \left (-\e^{\beta_j\mu_j} \right )\, , 
\label{ecurr2}
\end{equation} 
where $\li_s$ is the polylogarithm function. These expressions are the building blocks for the parameters 
$\dQ$ and $\dS$ and the efficiency coefficients $\eta$ and $\etat$. They depend on $(\beta_i,\mu_i)$ 
and $|\S_{ij}|^2$, which (due to the unitarity of $\S$) leads to $n(n+3)/2$ independent 
real parameters. We observe in particular that  
the currents depend on $\mu_i$ separately and not on differences $\mu_i-\mu_j$. 
In order to reduce this large number 
it is instructive to start by considering the system with two leads shown in Fig. \ref{junction2}.  

\bigskip 

\section{The case of two leads} 
\subsection{The LB state}
\medskip 

In the case $n=2$ there is only one transmission probability $|\S_{12}|^2=|\S_{21}|^2$ and 
the currents (\ref{ccurr2},\ref{ecurr2}) take the form  
\begin{equation}
J^N_1 = -J^N_2= 
\frac{|\S_{12}|^2}{2\pi \beta_1} 
\left [\ln \left (1+\e^{-\lambda_1} \right ) - 
r \ln \left (1+\e^{-\lambda_2} \right )\right ]\, , 
\label{ccurr4}
\end{equation} 
\begin{equation}
J^E_1  = -J^E_2 = 
-\frac{|\S_{12}|^2}{2\pi \beta_1^2} 
\left [\li_2 \left (-\e^{-\lambda_1} \right ) - 
r^2 \li_2 \left (-\e^{-\lambda_2} \right ) \right ] \, , 
\label{ecurr4}
\end{equation} 
where 
\begin{equation} 
r=\frac{\beta_1}{\beta_2}\, , \qquad \lambda_i = -\beta_i \mu_i \, , \quad i=1,2\, .    
\label{not1}
\end{equation} 
The factorization of $|\S_{12}|^2$ implies that the signs of $\dQ$ and $\dS$ depend 
exclusively on the parameters of the two heat reservoirs $R_{1,2}$, which greatly simplifies the 
study of the energy transmutation in the junction. 

We focus first on $\dQ$ denoting by $\D_\mp$ the 
domains in the space of parameters $(\beta_i, \mu_i)$ where $\dQ\lessgtr 0$ respectively. 
Without loss of generality one can assume $0\leq r \leq 1$. The direct investigation of 
\begin{equation}
\dot Q (\lambda_1,\lambda_2,r) =  
\frac{|\S_{12}|^2}{2\pi \beta^2_1} 
(\lambda_1 - r \lambda_2) 
\left [r \ln \left (1+\e^{-\lambda_2} \right )-\ln \left (1+\e^{-\lambda_1} \right )\right ] \, , 
\label{hf2}
\end{equation} 
shows that 
\begin{equation}
\D_- =  D_1 \cup D_2 \cup D_3 \, , 
\label{dom1}
\end{equation}
with   
\begin{eqnarray} 
D_1 &=& \{0<\lambda_1 \leq \lambda_2\, ,\, 0\leq r < r_1\}\, , 
\label{domains1} \\
D_2 &=& \{\lambda_1 > \lambda_2\, ,\, \lambda_1>0 \, ,\, 0\leq r < r_2\}\, ,
\label{domains2} \\ 
D_3 &=& \{0\geq \lambda_1 > \lambda_2 \, ,\, r_1< r < r_2\}\, , 
\label{domains3}
\end{eqnarray}
where 
\begin{equation} 
r_1 = \frac{\lambda_1}{\lambda_2}\, , \qquad r_2 = 
\frac{\ln \left (1+\e^{-\lambda_1} \right )}{\ln \left (1+\e^{-\lambda_2} \right )}\, .  
\label{r12}
\end{equation} 
The domain $\D_-$ has a complicated structure due to the dependence of $r_{1,2}$ on $\lambda_i$. 

We are ready at this point to compute the efficiency $\eta$. 
Using that $J^Q_1>0$ and $J_2^Q<0$ in $\D_-$, one gets from (\ref{eta1}) 
\begin{equation} 
\eta(\lambda_1,\lambda_2;r) = \frac{(\lambda_1-r\lambda_2)
\left [\ln \left (1+\e^{-\lambda_1} \right ) - 
r \ln \left (1+\e^{-\lambda_2} \right )\right ]}
{\lambda_1\left [\ln \left (1+\e^{-\lambda_1} \right ) - 
r \ln \left (1+\e^{-\lambda_2} \right )\right ]-\left [\li_2 \left (-\e^{-\lambda_1} \right ) - 
r^2 \li_2 \left (-\e^{-\lambda_2} \right ) \right ]}\, . 
\label{eta2}
\end{equation} 
This is an exact and explicit result for the efficiency of the Schr\"odinger junction 
in transforming heat to chemical potential energy at criticality. The expression (\ref{eta2}) shows 
the power of scale invariance and makes evident the advantage of the above approach 
with respect to the linear response approximation, which gives information about 
(\ref{eta2}) only in the neighborhood of $\lambda_1 \sim \lambda_2$ and $r\sim 1$. 
We demonstrate in appendix A that in this neighborhood the efficiency (\ref{eta2}) 
reproduces exactly the result of the linear response theory in \cite{bcps}.  

The analysis of (\ref{eta2}) in $\D_-$ shows that the maximal efficiency is obtained in the 
limit $\lambda_1=\lambda_2\equiv \lambda \to +\infty$. In fact one has 
\begin{equation}
\eta_{\rm max}(r) = \lim_{\lambda \to +\infty} \eta (\lambda,\lambda; r) = 1-r \equiv \eta_C\, , 
\label{eta3}
\end{equation}
which is the well known Carnot efficiency. According to (\ref{hf2}), in this limit the heat energy conversion vanishes 
$\lim_{\lambda \to \infty} \dQ(\lambda,\lambda,r)=0$.  

\begin{figure}[h]
\begin{center}
\begin{picture}(260,100)(80,25) 
\includegraphics[scale=0.75]{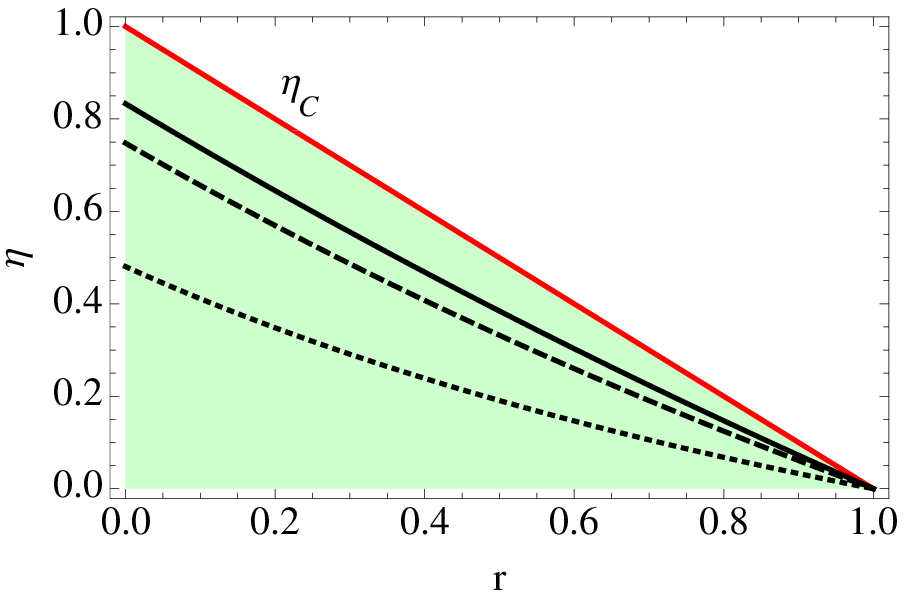} 
\quad 
\includegraphics[scale=0.75]{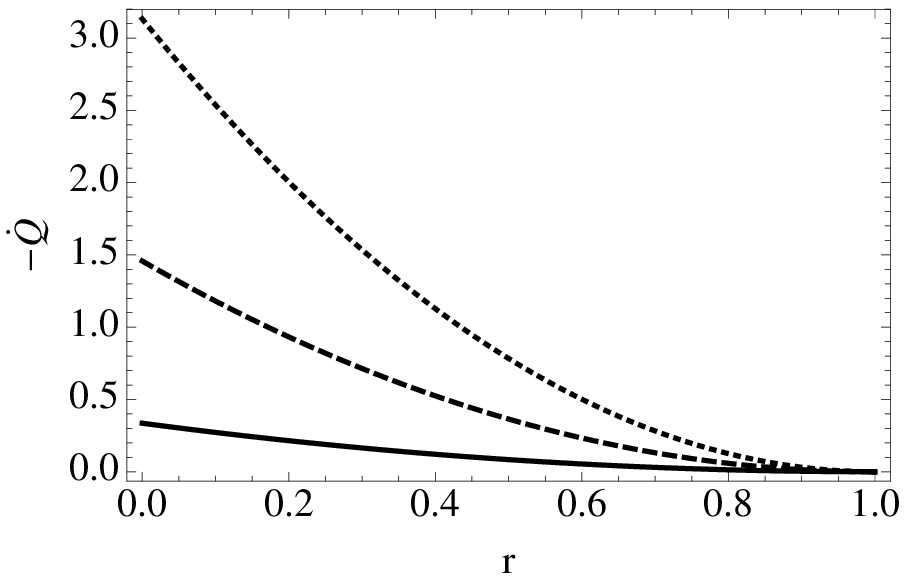} 
\end{picture} 
\end{center}
\caption{The efficiency $\eta$ compared to $\eta_C$ (left) and the heat $-\dQ$ converted to chemical energy (right) 
with $\lambda=1$ (dotted), $\lambda=3$ (dashed) and $\lambda=5$ (continuous) .} 
\label{plot1}
\end{figure}

Fig. \ref{plot1} illustrates the behavior of $\eta$ and the heat converted to chemical energy 
$-\dQ$ for some values of the parameter $\lambda\equiv \lambda_1=\lambda_2$. The plots in this figure 
show that the chemical energy production decreases with increasing the efficiency. 

\begin{figure}[h]
\begin{center}
\begin{picture}(80,100)(80,25) 
\includegraphics[scale=0.75]{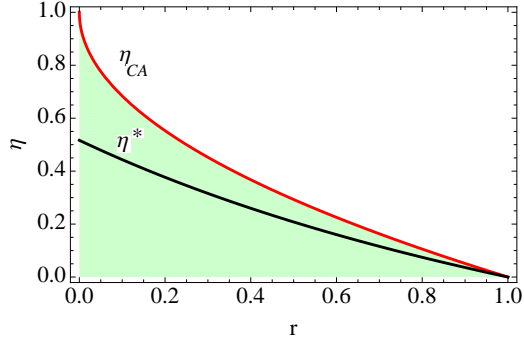}
\end{picture} 
\end{center}
\caption{The efficiency $\eta^*$ and the Curzon-Ahlborn bound $\eta_{CA}$.} 
\label{plot2}
\end{figure}

Another physically interesting regime is obtained by maximizing the chemical energy production 
$-\dQ(\lambda_1,\lambda_2,r)$ with respect to $\lambda_i$. 
A simple analysis shows that this function reaches its maximum 
at $\lambda_1=\lambda_2 \equiv \lambda^*$, where 
\begin{equation}
\lambda^* - (1+\e^{\lambda^*}) \ln (1+\e^{-\lambda^*}) =0\, . 
\label{sol1}
\end{equation} 
The solution of (\ref{sol1}) is $\lambda^*=1.14455...$ and the efficiency (\ref{eta2}) takes the form 
\begin{equation} 
\eta^*(r) \equiv \eta(\lambda^*,\lambda^*;r) = \frac{(1-r) \lambda^* \ln \left (1+\e^{-\lambda^*} \right )}
{\lambda^* \ln \left (1+\e^{-\lambda^*} \right ) - (1+r) \li_2 \left (-\e^{-\lambda^*} \right )}\, . 
\label{eta4}
\end{equation}
The quantity $\eta^*(r)$ 
is the counterpart of the concept of efficiency at maximal power, used in the context of heat engines. The plot 
in Fig. 4 shows that (\ref{eta4}) satisfies the Curzon-Ahlborn bound 
\begin{equation}
\eta^*(r)\leq 1-\sqrt r \equiv \eta_{CA}\, , 
\label{CAb}
\end{equation}
for all $r\in [0,1]$, which was known previously from linear response theory only for $r\sim 1$. We 
recall that this bound has been proposed for heat engines in the framework 
of endoreversible thermodynamics in \cite{ca}. The rigorous proof \cite{vb} covers the 
linear response regime, but it is known \cite{tu-08,bl-12,sb-12} that away of this regime 
the bound is not universal and can be violated. The possibility to enhance $\eta^*$ above $\eta_{CA}$ in 
our context is discussed in section 5.3. 

We turn now to the case when chemical energy is transformed in heat, namely 
\begin{equation}
\dQ = \mu_1 J_1^N + \mu_2 J_2^N = (\mu_1-\mu_2)J_1^N >0\, . 
\label{ch1}
\end{equation}
It is convenient to use in the domain $\D_+$ the coordinates $(\beta_i,\mu_i)$. Keeping $\beta_i$ 
arbitrary, one can assume without loss of generality that $\mu_1>\mu_2$ and set 
$\D_+=\{\beta_1,\beta_2,\mu_1>\mu_2\}$. From 
(\ref{ch1}) and the Kirchhoff rule one infers that $J_1^N > 0 > J_2^N$ on $\D_+$. Using 
this information one finds that the efficiency $\etat$, defined by (\ref{etat1}), can be expressed in terms 
of the parameter $u\equiv \mu_2/\mu_1$ in the simple form  
\begin{equation}
\etat (u) = 
\begin{cases} 
1-u\, ,   & \qquad  \mu_1 > \mu_2 \geq 0\, , \\
1\, ,   & \qquad  \mu_1 \geq 0 > \mu_2\, , \\
1-1/u\, ,   & \qquad  0>\mu_1 > \mu_2\, , \\
\end{cases}
\qquad \quad u=\frac{\mu_2}{\mu_1}\, , 
\label{etat2}
\end{equation}
fully covering the domain $\D_+$. The formula (\ref{etat2}) describes in exact form the conversion 
of chemical potential energy in heat. For $n=2$ the efficiency $\etat$ does not depend on the temperatures 
and the explicit form of the currents $\mu_i J_i^N$, but 
only on the values of the chemical potentials $\mu_i$. 
The first line of (\ref{etat2}) resembles the Carnot formula, where the temperature ratio $r$ is 
substituted by the chemical potential ratio $u$. The third line instead takes into account that differently 
from the temperatures, the chemical potentials can take also negative values. Finally, 
we observe that for $\mu_1 > 0 > \mu_2$ both chemical potential currents 
$\mu_i J_i^N$ are flowing towards the junction, 
transforming the chemical energy in heat completely ($\etat =1$). Since in this case 
the domains of $\mu_i$ are separated by the point $\mu=0$, this regime of 
energy transmutation in the junction can not be reached by linear response theory. 

\begin{figure}[h]
\begin{center}
\begin{picture}(80,100)(80,25) 
\includegraphics[scale=0.75]{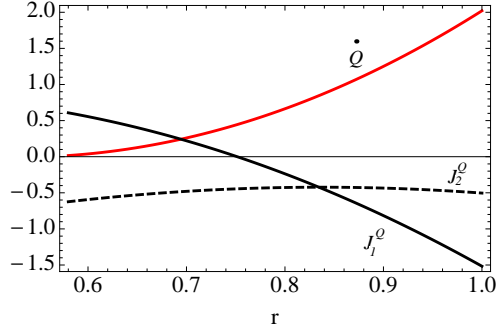}
\end{picture} 
\end{center}
\caption{The flow $\dQ>0$ and the currents $J_1^Q$ and $J_2^Q$ for $\lambda_1=-1.5$ and $\lambda_2=-3$.}  
\label{fig5}
\end{figure}

We have seen that for $\dQ<0$ the junction can be compared to a heat engine. In analogy, 
one can attempt to interpret the system for $\dQ>0$ as a refrigerator or a heat pump. 
A careful analysis shows that this possible only partially because there are subsets in $\D_+$ 
where both heat currents have the same sign, i.e. they are both leaving or entering the 
heat reservoirs. A typical situation is shown in Fig. \ref{fig5}, where $J_1^Q$ and $J_2^Q$ 
become both negative (i.e. entering the heat reservoirs) for $r>0.75$, which 
is not the case of conventional refrigerators and heat pumps. For this reason the 
standard coefficients of performance for a refrigerator and a heat pump can not be applied in the whole 
domain $\D_+$. As already mentioned, this is our main motivation to introduce the efficiency 
$\etat$ by (\ref{etat1}), which has the advantage of working everywhere in $\D_+$. 

Let us consider finally the entropy production. Substituting (\ref{ccurr4},\ref{ecurr4}) in  (\ref{entr1}) one obtains 
\begin{eqnarray}
\dS = \frac{|\S_{12}|^2}{2\pi r\beta_1} \Bigl \{(1-r) \left [ 
r^2 \li_2 \left (-\e^{-\lambda_2} \right ) - \li_2 \left (-\e^{-\lambda_1} \right ) \right ] 
\nonumber \\
+\, r (\lambda_1-\lambda_2) 
\left [r \ln \left (1+\e^{-\lambda_2} \right )-\ln \left (1+\e^{-\lambda_1} \right ) \right ]\Bigr \} > 0\, ,
\label{entr2}
\end{eqnarray} 
confirming the general statement \cite{gn} about the entropy production in the LB state. 

\bigskip

\subsection{Transport in the orbit $O_{\beta,\mu}$}
\medskip

We consider here the quantum transport in the states $\{\Omega^X_{\beta,\mu}\, :\, X=P,T,PT\}\subset \H_{\rm LB}$. 
Using (\ref{pt1}), the current expectation values in these states are simply expressed in terms of (\ref{ccurr1},\ref{ecurr1}) 
as follows: 
\begin{eqnarray}
P\, &:&\, J_i^N \longmapsto -J_i^N\, \qquad J_i^E \longmapsto -J_i^E\, , 
\label{P}\\
T\, &:&\, J_i^N \longmapsto J_i^N\, \qquad \; \; \; J_i^E \longmapsto -J_i^E\, , 
\label{T}\\ 
P\,T\, &:&\, J_i^N \longmapsto -J_i^N\, \qquad J_i^E \longmapsto J_i^E\, . 
\label{PT}
\end{eqnarray}
Let us observe in passing that the action of the charge conjugation 
\begin{equation}
C\, : \, J_i^N \longmapsto -J_i^N\, \qquad J_i^E \longmapsto J_i^E\, ,  
\label{C}
\end{equation}
coincides with the $PT$ operation (\ref{PT}). In fact, $CPT$ on the above currents 
is the identity transformation. 

The minus sign appearing in some currents affects the quantum transport. Moreover, the entropy 
production $\dS^X$ in the state $\Omega^X_{\beta,\mu}$ differs from that (\ref{entr2}) in 
the LB state. In fact, combining (\ref{entr1}) and (\ref{P}) we conclude that 
$\dS^P \leq 0$ for all values of the heat reservoir parameters $(\beta_i,\mu_i)$. We see that the state 
$\Omega^P_{\beta,\mu}$, which from the microscopic point of view is a well defined state of the system, 
violates the second law of thermodynamics and therefore has no admissible macroscopic behavior. 
The situation with the states $\Omega^T_{\beta,\mu}$ and $\Omega^{PT}_{\beta,\mu}$ is more subtle. 
The analysis shows that for these two states there exist domains 
in the whole parameter space $(\beta_i,\mu_i)$, where $\dS^T>0$ and $\dS^{PT}>0$. In these domains 
the quantum transport is consistent with the laws of thermodynamics. We consider for illustration the family of states 
$\{\Omega^{PT}_{\beta,\mu}\, :\, \lambda_1=\lambda_2\equiv \lambda\}$. The entropy production there is 
\begin{equation}
\dS^{PT} = -\frac{|\S_{12}|^2}{2\pi r\beta_1} (1-r)^2(1+r) \li_2 \left (-\e^{-\lambda} \right ) > 0\, . 
\label{entr3}
\end{equation} 
Moreover 
\begin{equation}
\dQ^{PT}(\lambda,\lambda,r)= \frac{|\S_{12}|^2}{2\pi \beta^2_1}
(1-r)^2\lambda \ln \left (1+\e^{-\lambda} \right ) <0\, \qquad {\rm for}\quad \lambda <0\, , 
\label{qpt}
\end{equation} 
and the efficiency of transforming heat into chemical potential energy is given by 
\begin{equation} 
\eta^{PT}(\lambda,r) = \frac{(1-r) \lambda \ln \left (1+\e^{-\lambda} \right )}
{\lambda \ln \left (1+\e^{-\lambda} \right ) + (1+r) \li_2 \left (-\e^{-\lambda} \right )}\, , \qquad \lambda < 0\, .  
\label{eta5}
\end{equation} 
This result resembles (\ref{eta4}), but for a sign in the denumerator. 
The relative maximum is 
\begin{equation} 
\eta_{\rm max}^{PT}(r)= \lim_{\lambda \to -\infty}  \eta^{PT}(\lambda,r) = \frac{2(1-r)}{r+3} \not=\eta_C \, , 
\label{eta6}
\end{equation}
which coincides also with the efficiency at maximal chemical energy production 
\begin{equation} 
\eta^{*PT}(r)= \frac{2(1-r)}{r+3} <  \eta_{CA}\, . 
\label{eta6star}
\end{equation} 
These results show that the efficiency in the state $\Omega_{\beta,\mu}^{PT}$ 
differs from that in $\Omega_{\beta,\mu}$. We will elaborate more on the value of $\eta^{*PT}$ few lines below. 

Concerning the regime $\dQ>0$, it is easy to deduce from (\ref{P}-\ref{PT}) that all four states in $O_{\beta,\mu}$ 
have the same efficiency $\etat$ given by (\ref{etat2}). 

Summarizing, parity and time reversal have an important impact on the quantum transport and 
efficiency. Indeed, we have shown that there are regions in the parameter space of the states 
$\Omega^T_{\beta,\mu}$ and $\Omega^{PT}_{\beta,\mu}$, where the second law of 
thermodynamics is satisfied and the efficiency $\eta$ has a physically acceptable value, which 
differs from that in the LB state $\Omega_{\beta,\mu}$.  

\bigskip 

\subsection{Comments about $\eta^*$} 
\medskip 

It is instructive to compare now (\ref{eta4},\ref{eta6star}) with some exact results about the efficiency 
at maximal power obtained for other systems. Applying stochastic thermodynamics to a simple 
model of classical particle transport, the following explicit expression 
\begin{equation}
\eta_{cp}^*=\dfrac{\eta_C^2}{\eta_C-(1-\eta_C)\ln(1-\eta_C)} 
\label{bl}
\end{equation}
has been derived in \cite{bl-12}. The same expression has been obtained \cite{tu-08} for the Feynman's ratchet model 
as a heat engine. For a Brownian particle undergoing a Carnot cycle it has been found \cite{ss-08} that 
\begin{equation}
\eta_{Bp}^*=\dfrac{2\eta_C}{4-\eta_C}\, , 
\label{ss}
\end{equation} 
which, remarkably enough, coincides with the Schr\"odinger junction 
efficiency $\eta^{*PT}$ given by (\ref{eta6star}). 
The case of electron transport through a quantum dot has 
been treated in \cite{ebl-09,ssj}. 

It has been observed in \cite{tu-08,bl-12,ssj} that away ($r<0.5$) from the linear response regime, 
the efficiency (\ref{bl}) exceeds the 
Curzon-Ahlborn bound. One can wonder if a mechanism exists to enhance the efficiency (\ref{eta4}) 
in the LB state $\Omega_{\beta,\mu}$ above $\eta_{CA}$ as well. This question attracted 
recently some attention \cite{bcps}, the proposal being to couple the system with an appropriate external potential. 
In \cite{bcps} the effect of a classical magnetic field has been explored in the linear response approximation. 
We describe here an alternative, which simplifies our previous construction in \cite{mss-13} both from the technical 
and physical points of view. The main idea is   
based on the fact that in our general setting the heat reservoir 
dispersion relations $\omega_i$ need not to be equal and/or to coincide with the bulk dispersion 
relation $\omega$. So, let us perform the shift $\omega \longmapsto \omega -v$, which 
is equivalent to the introduction of a constant potential $V=-v$ in the bulk equation of motion (\ref{eqm1}). 
If one performs the same shift in $\omega_i$ the efficiency will not change. One can imagine however 
to screen the reservoirs $R_i$ from the potential $V$, thus keeping $\omega_i = k^2/2m$ invariant. 
This operation does not affect the particle currents $J_i^N$ which, according to (\ref{ccurr1}), 
depend only on $\omega_i$. From (\ref{ecurr1}) one infers however that the shift 
in $\omega$ modifies the energy transport in the following simple way 
\begin{equation}
J_i^E \longmapsto J_i^E - v J_i^N\, . 
\label{shift1}
\end{equation}
In the domain $\D_-$, which is still given by (\ref{dom1}), one finds 
the efficiency  
\begin{equation} 
\eta(\lambda_1,\lambda_2;r,a) = \frac{(\lambda_1-r\lambda_2)
\left [\ln \left (1+\e^{-\lambda_1} \right ) - 
r \ln \left (1+\e^{-\lambda_2} \right )\right ]}
{(\lambda_1-a)\left [\ln \left (1+\e^{-\lambda_1} \right ) - 
r \ln \left (1+\e^{-\lambda_2} \right )\right ]-\left [\li_2 \left (-\e^{-\lambda_1} \right ) - 
r^2 \li_2 \left (-\e^{-\lambda_2} \right ) \right ]}\, , 
\label{eta2a}
\end{equation} 
with $a\equiv \beta_1 v$ being dimensionless. 
The maximal efficiency is attained at $\lambda_1=\lambda_2 \to +\infty$ and equals $\eta_C$ as before. For 
$\eta^*$ one finds instead 
\begin{equation} 
\eta^*(r,a) \equiv \eta(\lambda^*,\lambda^*;r,a) = \frac{(1-r) \lambda^* \ln \left (1+\e^{-\lambda^*} \right )}
{(\lambda^*-a) \ln \left (1+\e^{-\lambda^*} \right ) - (1+r) \li_2 \left (-\e^{-\lambda^*} \right )}\, , 
\label{eta4a}
\end{equation} 
which reproduces (\ref{eta4}) for $a=0$. For the moment $a$ is a free real parameter, 
but the condition (\ref{nentr3}) is violated and the sign 
of $\dS$ needs to be investigated. Imposing $\dS>0$ for all $r\in [0,1]$, one obtains the constraint 
\begin{equation}
a< \frac{-\li_2 \left (-\e^{-\lambda^*} \right )}{\ln \left (1+\e^{-\lambda^*} \right )} = 1,07122... \equiv a^*\, . 
\label{enh1}
\end{equation} 

\begin{figure}[h]
\begin{center}
\begin{picture}(80,100)(80,25) 
\includegraphics[scale=0.75]{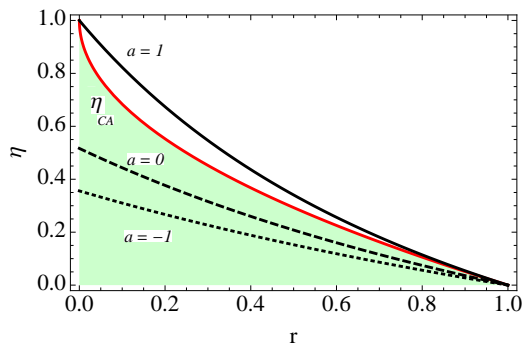}
\end{picture} 
\end{center}
\caption{The efficiency $\eta^*(r,a)$ for different values of $a$ compared to $\eta_{CA}$.} 
\label{fig.6}
\end{figure}

It is easy to deduce from (\ref{eta4a}) that for $a\in (0,a^*)$ the efficiency $\eta^*(r,a)$ 
is enhanced above $\eta^*(r,0)=\eta^*(r)$ given by (\ref{eta4}). The opposite effect is 
observed for $a\in (-\infty ,0)$. This behavior 
of the efficiency is explained by the following intuitive physical argument. 
Since $\omega_i-\omega=a/\beta_1$, positive values of $a$ favor the particle emission from the heat reservoirs 
$R_i$ to the leads $L_i$, thus improving the efficiency. Negative values of $a$ are instead 
damping this emission, causing the opposite effect. 

Fig. \ref{fig.6} illustrates the basic properties of $\eta^*(r,a)$. In particular, 
we see that for $a\sim 1$ the Curzon-Ahlborn bound is exceeded. 

Finally, concerning the experimental realization of the above theoretical setup, 
one possibility could be to use the electric field produced by a wire with constant linear static charge 
distribution, which is located parallel at a finite distance from the two-terminal devise in Fig. \ref{junction2}. 
This charged wire produces a constant electric field along the devise. 
When the heat baths are screened by metallic boxes, the electric field provides 
the constant shift needed in the bulk dispersion relation.

\bigskip

\section{Junctions with three leads}

\medskip 

The new element in the treatment of the case with $n>2$ leads is that the transmission amplitudes 
$|\S_{ij}|^2$ do no longer factorize in front of the currents. For $n=3$  the transport properties depend on 9 
parameters. The linear response approximation has been studied recently in \cite{sns-14,ewa}. Here 
we pursue further the exact analysis of the scale invariant case for  
\begin{equation}
\beta_1 \mu_1 = \beta_2 \mu_2 = \beta_3 \mu_3 \equiv -\lambda \, , \qquad 
r=\frac{\beta_1}{\beta_2}\, , \qquad s=\frac{\beta_1}{\beta_3}\, . 
\label{fug}
\end{equation}
Using (\ref{symS}), one obtains from (\ref{ccurr2},\ref{ecurr2}) the heat currents 
\begin{eqnarray}
J^Q_1 = 
\frac{1}{2\pi \beta^2_1} \Bigl \{ 
\lambda \ln \left (1+\e^{-\lambda } \right )\left [ |\S_{12}|^2 (1-r) +|\S_{13}|^2 (1-s) \right ]
\nonumber \\
- \li_2 \left (-\e^{-\lambda} \right ) 
\left [ |\S_{12}|^2 (1-r^2) +|\S_{13}|^2 (1-s^2) \right ]  \Bigr \} \, ,
\label{TJ1}
\end{eqnarray}
\begin{eqnarray}
J^Q_2 = 
\frac{1}{2\pi \beta^2_1} \Bigl \{ 
r \lambda \ln \left (1+\e^{-\lambda } \right )\left [ |\S_{23}|^2 (r-s) +|\S_{12}|^2 (r-1) \right ]
\nonumber \\
- \li_2 \left (-\e^{-\lambda} \right ) 
\left [ |\S_{23}|^2 (r^2-s^2) +|\S_{12}|^2 (r^2-1) \right ]  \Bigr \} \, ,
\label{TJ2}
\end{eqnarray}
\begin{eqnarray}
J^Q_3 = 
\frac{1}{2\pi \beta^2_1} \Bigl \{ 
s \lambda \ln \left (1+\e^{-\lambda } \right )\left [ |\S_{23}|^2 (s-r) +|\S_{13}|^2 (s-1) \right ]
\nonumber \\
- \li_2 \left (-\e^{-\lambda} \right ) 
\left [ |\S_{23}|^2 (s^2-r^2) +|\S_{13}|^2 (s^2-1) \right ]  \Bigr \} \, .
\label{TJ3}
\end{eqnarray}
For the entropy production and heat flow on gets 
\begin{equation}
\dS = -\frac{\li_2 \left (-\e^{-\lambda} \right )}{2\pi s r\beta_1} 
\left [|\S_{12}|^2 s (1+r)(1-r)^2 + |\S_{13}|^2 r (1+s)(1-s)^2  + |\S_{23}|^2(r+s)(r-s)^2 \right ] > 0\, ,  
\label{3entr3}
\end{equation} 
\begin{equation}
\dQ = -\frac{1}{2\pi \beta^2_1}
\lambda \ln \left (1+\e^{-\lambda} \right ) \left [|\S_{12}|^2 (1-r)^2 +|\S_{13}|^2 (1-s)^2 + |\S_{23}|^2 (r-s)^2 \right ] \, . 
\label{qpt3}
\end{equation} 
As expected, the entropy production in the LB state is always positive. For $\lambda > 0$ one has $\dQ<0$ and 
one can study the efficiency (\ref{eta1}) of transforming heat to chemical potential energy. 
In order to do that one has to determine first the sign of the currents (\ref{TJ1}-\ref{TJ3}). Without 
loss of generality one can assume that 
\begin{equation}
r<1\, , \qquad s<1\, ,\qquad r<s\, , 
\label{cond1} 
\end{equation} 
which implies the following ordering $T_2<T_3<T_1$ among the temperatures of the heat reservoirs. 
For simplifying the analysis we assume to end of this section that $|\S_{13}|^2=0$, devoting 
the appendix B to the case of generic transmission amplitudes. Then (\ref{TJ1}-\ref{TJ3}) imply 
$J^Q_1>0,\, J^Q_2 <0,\, J^Q_3 >0$ for $\lambda >0$. With this information one obtains from (\ref{eta1}) 
\begin{eqnarray}
\eta (\lambda;r,s) = \qquad  \qquad \qquad \qquad  \qquad  \qquad  \qquad  \qquad  \qquad  
\nonumber\\
\frac{\lambda \ln \left (1+\e^{-\lambda} \right )\left[|\S_{12}|^2 (1-r)^2 + |\S_{23}|^2 (s-r)^2\right ]}
{\lambda \ln \left (1+\e^{-\lambda} \right ) \left[|\S_{12}|^2 (1-r) + |\S_{23}|^2 s (s-r)\right ] - 
\li_2 \left (-\e^{-\lambda} \right )\left[|\S_{12}|^2 (1-r^2) + |\S_{23}|^2 (s^2-r^2)\right ]}\, .  
\nonumber \\
\label{eta7}
\end{eqnarray} 
The relative maximum is 
\begin{equation}
\eta_{\rm max}(r,s) = \lim_{\lambda \to +\infty} \eta (\lambda;r,s) = 
1- r \frac{|\S_{12}|^2 (1-r) + |\S_{23}|^2 (s-r)}{|\S_{12}|^2 (1-r) + |\S_{23}|^2 s (s-r)}\, , 
\label{eta8}
\end{equation}
which suggests to introduce an effective $r$-parameter 
\begin{equation}
r_{\rm eff}(s) =  r \frac{|\S_{12}|^2 (1-r) + |\S_{23}|^2 (s-r)}{|\S_{12}|^2 (1-r) + |\S_{23}|^2 s (s-r)}  \leq 1  
\qquad {\rm for} \quad r<1\, , r<s\, 
\label{eff}
\end{equation}
with the following physical meaning. The hotter reservoirs of our system are $R_1$ and $R_3$ because 
$T_1>T_2$ and $T_3>T_2$. Both $R_1$ and $R_3$ communicate with the cold reservoir 
$R_2$ via $\S_{12}$ and $\S_{23}$. 
Suppose now we replace $R_1$ and $R_3$ with one heat bath $R^\prime$ 
and ask about its temperature $T^\prime$, which gives the same efficiency. It turns out that 
the answer is $T^\prime=T_2/r_{\rm eff}$. In fact, one can express the 3-lead 
efficiency (\ref{eta7}) in terms of the 2-lead formula (\ref{eta2}) simply as 
\begin{equation}
\eta (\lambda;r,s) = \eta(\lambda,\lambda;r_{\rm eff}(s)) \, .
\label{eff1}
\end{equation} 

We turn now to the regime of heat production $\dQ>0$ which, according to (\ref{qpt3}), 
takes place for $\lambda<0$. The corresponding efficiency is defined by (\ref{etat1}). 
Let us introduce the 
parameters 
\begin{equation}
u = \frac{\mu_2}{\mu_1}\, , \qquad v = \frac{\mu_3}{\mu_1}\, .
\label{chem}
\end{equation}
Because of (\ref{fug}), the conditions (\ref{cond1}) can be rewritten in the form 
\begin{equation} 
u<1\, , \qquad v<1\, ,\qquad u<v\, . 
\label{cond2} 
\end{equation} 
One can show now that $\mu_1J^N_1>0,\, \mu_2 J^N_2 <0,\, \mu_3J^N_3 >0$. 
Therefore, using (\ref{etat1}) one obtains 
\begin{equation}
\etat (u,v) = 
1- u \frac{|\S_{12}|^2 (1-u) + |\S_{23}|^2 (v-u)}{|\S_{12}|^2 (1-v) + |\S_{23}|^2 v (v-u)}\, . 
\label{etat8}
\end{equation}
The effective $u$-parameter now reads 
\begin{equation}
u_{\rm eff}(v) =  u \frac{|\S_{12}|^2 (1-u) + |\S_{23}|^2 (v-u)}{|\S_{12}|^2 (1-v) + |\S_{23}|^2 v (v-u)}  \leq 1  
\qquad {\rm for} \quad u<1\, , u<v\, 
\label{eff3}
\end{equation}
and 
\begin{equation}
\etat (u,v) = \etat(u_{\rm eff}(v)) \, ,
\label{eff4}
\end{equation} 
whose right hand side is given by the first line (since $u<1$) of the two-lead expression (\ref{etat2}). 

We refer to appendix B for the analysis of the general case, in which all three transmission 
amplitudes are nontrivial.

\bigskip 
\section{Bosonic junctions}
\medskip 

We illustrate in this section the influence of the statistics on the transport and efficiency 
of the Schr\"odinger junction. Substituting in (\ref{ccurr1},\ref{ecurr1}) the Fermi distribution $d_j(k)$ 
with the Bose one $b_j(k)$, the corresponding integrands develop a 
singularity at $k^2=2m \mu_i$. This singularity signals condensation like phenomena, whose 
consideration is beyond the scope of the present paper. For this reason we assume in this section 
$\mu_i < 0$. Focussing on the case 
$n=2$, the bosonic counterparts of (\ref{ccurr4},\ref{ecurr4},\ref{hf2},\ref{entr2}) are 
\begin{equation}
J_1^N = - J_2^N = 
\frac{\vert \mathbb{S}_{12}\vert^2}{2\pi\beta_1}\left[-\ln(1-\e^{-\lambda_1})+ r\ln(1-\e^{-\lambda_2})\right]\, , 
\label{JNbos2}
\end{equation}
\begin{equation}
J_1^E = - J_2^E = 
\frac{\vert \mathbb{S}_{12}\vert^2}{2\pi\beta_1^2}\left[\mathrm{Li}_2(\e^{-\lambda_1})- 
r^2\mathrm{Li}_2(\e^{-\lambda_2})\right]\, ,  
\label{JEbos2}
\end{equation}
\begin{equation}
\dQ^{(b)}(\lambda_1,\lambda_2, r) = \frac{\vert \mathbb{S}_{12}\vert^2}{2\pi\beta_1^2}
(\lambda_1-r \lambda_2) \left[\ln(1-\e^{-\lambda_1})- r\ln(1-\e^{-\lambda_2})\right] \, , 
\label{Qbos}
\end{equation} 
\begin{equation}
\dot{S}^{(b)}=\frac{\vert \mathbb{S}_{12}\vert^2}{2\pi r \beta_1} \left\lbrace (1-r) 
\left[\mathrm{Li}_2(\e^{-\lambda_1})- r^2\mathrm{Li}_2(\e^{-\lambda_2})\right] 
+ r(\lambda_1 - \lambda_2) \left[\ln(1-\e^{-\lambda_1})- r\ln(1-\e^{-\lambda_2})\right] \right\rbrace \, ,
\label{EntropyBos}
\end{equation}
where one should keep in mind that $\lambda_i \equiv -\beta_i\mu_i > 0$. The apex $(b)$ 
in (\ref{Qbos},\ref{EntropyBos}) and below is added in order to distinct the bosonic from the fermionic expressions. 
The domains $\D^{(b)}_\mp$, where $\dQ^{(b)}\lessgtr 0$, 
can be determined like in fermion case. 
One has 
\begin{equation}
\mathcal{D}^{(b)}_- = D^{(b)}_1\cup D^{(b)}_2
\label{dombos}
\end{equation}
with
\begin{gather}
D^{(b)}_1 = \lbrace 0 < \lambda_1 \leq \lambda_2 , \  0\leq r < r_1  \rbrace \, ,
\label{D1bos} \\
D^{(b)}_2 = \lbrace 0 < \lambda_2 < \lambda_1 , \ 0 \leq r < r_2  \rbrace \, ,
\label{D2bos}
\end{gather} 
\begin{equation}
r_1 = \dfrac{\lambda_1}{\lambda_2}\, , \ \ \
r_2= \dfrac{\ln(1-\e^{-\lambda_1})}{\ln(1-\e^{-\lambda_2})}\, .
\label{r12bos}
\end{equation} 
In $\D^{(b)}_-$ one finds the bosonic efficiency 
\begin{equation}
\eta^{(b)}(\lambda_1,\lambda_2, r)=
\frac{(\lambda_1-r \lambda_2) \left[ 
r\ln(1-\e^{-\lambda_2})-\ln(1-\e^{-\lambda_1})\right]}{\lambda_1 \left[ 
r\ln(1-\e^{-\lambda_2})-\ln(1-\e^{-\lambda_1})\right]+ 
\left[ \mathrm{Li}_2(\e^{-\lambda_1})- r^2\mathrm{Li}_2(\e^{-\lambda_2})\right]}\, .
\label{EffBos}
\end{equation} 
In the overlap of $\D_-$ and $\D^{(b)}_-$ one can compare 
(\ref{EffBos}) to the fermion efficiency (\ref{eta2}). For the same values of $\lambda_i$ in 
$D_1^{(b)}=D_1$ one finds that the bosonic efficiency exceeds the fermionic one. The same holds 
for the amount of heat transformed in chemical energy. This feature persists in the overlap 
of $D_2^{(b)}$ with $D_2$, except in the neighborhood of $r_2$, where one has the opposite behavior. 
The maximal bosonic efficiency is obtained for $\lambda_1=\lambda_2\equiv\lambda\rightarrow +\infty$ 
and coincides with the Carnot efficiency, 
\begin{equation}
\eta^{(b)}_{\rm{max}}(r)=\lim_{\lambda\rightarrow +\infty}\eta^{(b)}(\lambda,\lambda, r)=
1-r \equiv \eta_C\, .
\label{EffMaxBos}
\end{equation}

\begin{figure}[h]
\begin{center}
\begin{picture}(80,100)(80,25) 
\includegraphics[scale=0.75]{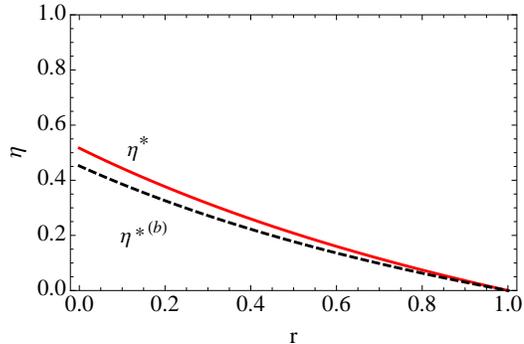}
\end{picture} 
\end{center}
\caption{Comparing the bosonic and fermionic efficiencies $\eta^{*(b)}$ and $\eta^*$.} 
\label{plot3}
\end{figure}

Let us derive now the bosonic efficiency at maximal chemical energy production. 
The maximum of the function $-\dQ^{(b)}(\lambda_1,\lambda_2, r)$ is 
reached at $\lambda_1=\lambda_2\equiv \lambda_b^*$, where
\begin{equation}
\lambda_b^* - (1-\e^{\lambda_b^*})\ln(1-\e^{-\lambda_b^*})=0\, .
\label{MaxQbos}
\end{equation}
The solution of (\ref{MaxQbos}) is $\lambda_b^* = 0.69314 \ldots$ and the efficiency reads
\begin{equation}
\eta^{*(b)} (r)\equiv \eta^{(b)}(\lambda_b^*,\lambda_b^*, r)=
\frac{(1-r)\lambda_b^* \ln \left (1-\e^{-\lambda_b^*}\right )}
{\lambda_b^* \ln\left (1-\e^{-\lambda_b^*}\right )-
(1+r) \mathrm{Li}_2\left (\e^{-\lambda_b^*}\right )}\, .
\label{EffQmaxBos}
\end{equation} 
The study of (\ref{EffQmaxBos}) reveals that at maximal chemical energy production 
the bosonic junction is slightly less efficient then the fermion one, as shown in Fig. \ref{plot3}. 
The enhancement mechanism for fermions, described in section 5.3, applies to 
bosonic junctions as well. 
 
We conclude by observing that in the regime $\dQ^{(b)}>0$ the bosonic efficiency 
$\etat^{(b)}$ of transforming chemical energy in heat precisely coincides with the fermionic one (\ref{etat2}).

\bigskip 
\section{Conclusions} 
\medskip 

We described in this paper a basic process of transformation of heat in 
chemical potential energy and vice versa, which takes place in systems away from equilibrium. 
The phenomenon is universal and stems from the fact that even if the total energy 
of the system is conserved, the heat and chemical potential energies are in general not separately conserved. 
Both directions of the process of energy transmutation are characterized by their own efficiency coefficient.  
The specific features of the phenomenon depend 
on the particle statistics and on the choice of nonequilibrium state. We illustrated this fact by studying 
the fermionic and bosonic Schr\"odinger junctions in the LB state and its orbit under 
parity and time reversal. The relative quantum transport depends in a complicated nonlinear way 
on the temperatures and chemical potentials, which parametrize the nonequilibrium states. In order 
to control these characteristics of the system, we avoided the use of any approximation and in particular, of the 
linear response theory. Assuming that the interaction which drives the system away from equilibrium  
is scale invariant, we described the quantum transport exactly and derived the 
explicit expressions of the efficiency coefficients. 

Exploring the orbit of the LB state under parity and time reversal transformations, we have shown that 
space-time symmetries have an essential impact on the quantum transport and efficiencies. 
Concerning the internal symmetries, our model has a global $U(1)$-symmetry associated 
with the particle number. 
For systems with a larger internal symmetry group, the process of energy transmutation 
becomes even more involved, due to the 
presence of several types of chemical potentials. In this case the Landauer-B\"uttiker state 
is induced not by a Gibbs state, but by a generalized Gibbs ensemble \cite{j-57} 
generated by a complete set of commuting charges of the extended symmetry group. 
The study of such nonequilibrium states may provide an important insight in the role of 
internal symmetries in quantum transport.

\bigskip

\leftline{\bf Acknowledgments:}
\medskip 

M.M. would like to thank the Laboratoire de Physique Th\'eorique d'Annecy-le-Vieux for the 
kind hospitality during the preparation of the manuscript.

\appendix
\bigskip 
\section{Contact with linear response theory} 
\medskip 

We show here that the exact efficiency (\ref{eta2}) reproduces in the linear response regime the result 
of \cite{bcps}. The meeting point of the two frameworks is the Onsager matrix $X$ \cite{call}. Setting 
\begin{equation}
\beta_1=\beta \, ,\quad \beta_2 = \beta + \delta \beta\, , \quad 
\mu_1=\mu\, , \quad  \mu_2 = \mu + \delta \mu\, ,  \quad  \lambda = - \beta \mu \, , 
\label{X0}
\end{equation}
the entries $X_{ij}$ of $X$ are defined by \cite{call} 
\begin{eqnarray} 
-J_1^N = X_{11}\, \beta\, \delta \mu + X_{12}\, \delta \beta + \cdots \, , \nonumber \\ 
J_1^Q =J_1^E -\mu J_1^N = X_{21}\, \beta\, \delta \mu + X_{22}\, \delta \beta + \cdots \, , 
\label{X1}
\end{eqnarray} 
where the dots stand for higher orders of the expansion in $\delta \mu$ and $\delta \beta$. 
For the Onsager matrix of our system with scale invariant interaction in the junction one gets from \cite{bcps} 
\begin{eqnarray} 
X_{11} &=& \frac{|\S_{12}|^2}{2\pi \beta} \frac{1}{1+\e^{\lambda}}\, , \nonumber \\ 
X_{12} &=& X_{21} = -\frac{|\S_{12}|^2}{2\pi \beta^2} 
\left [\frac{\lambda}{1+\e^{\lambda}}+ \ln \left (1+\e^{-\lambda} \right ) \right ] \, , \nonumber \\ 
X_{22} &=& \frac{|\S_{12}|^2}{2\pi \beta^2} 
\left [\frac{\lambda^2}{1+\e^{\lambda}}+ 2\lambda \ln \left (1+\e^{-\lambda} \right )- 
\li_2 \left (-\e^{-\lambda} \right )\right ] \, . 
\label{X2} 
\end{eqnarray} 
Moreover, according to \cite{bcps} the linear response efficiency $\eta_{{}_{\rm LR}}$ is given by 
\begin{equation} 
\eta_{{}_{\rm LR}} = \frac{-(X_{11}\, \beta\, \delta \mu + X_{12}\, \delta \beta)\, \delta \mu} 
{X_{21}\, \beta\, \delta \mu + X_{22}\, \delta \beta}  \, . 
\label{X3}
\end{equation} 
Inserting (\ref{X2}) in (\ref{X3}) and expanding in $\delta \mu$ and $\delta \beta$  
one obtains to the first order  
\begin{eqnarray} 
\eta_{{}_{\rm LR}} = \frac {\e^{-\lambda}}
{(1+\e^{-\lambda})\ln \left (1+\e^{-\lambda} \right ) +\lambda \e^{-\lambda}}\, \beta\, \delta \mu - 
\qquad \qquad \qquad \; \nonumber \\ 
\frac {(1+\e^{-\lambda})\ln^2 \left (1+\e^{-\lambda} \right ) +2 \e^{-\lambda}\li_2 \left (-\e^{-\lambda} \right )} 
{\beta [(1+\e^{-\lambda})\ln \left (1+\e^{-\lambda} \right ) +\lambda \e^{-\lambda}]^2}\, \delta \beta + \cdots \, . 
\label{X4}
\end{eqnarray} 
This result coincides precisely with the expansion of the exact efficiency (\ref{eta2}) (expressed in terms of  
the variables (\ref{X0})) to the first order in $\delta \mu$ and $\delta \beta$, which concludes the proof. 
The above argument has a direct generalization to the case $n>2$. 

\bigskip 

\section{The 3-lead junction with generic $\S$-matrix amplitudes} 

Combining (\ref{TJ1},\ref{TJ2}) with (\ref{cond1}), we conclude that $J_1^Q>0$ and $J_2^Q<0$. 
So, one is left with the study of the sign of $J_3^Q$ given by (\ref{TJ3}). 
The coefficients of the logarithm and the polylogarithm are both positive when 
\begin{equation} 
s >  \dfrac{r\vert \mathbb{S}_{23}\vert^2 + \vert \mathbb{S}_{13}\vert^2}
{\vert \mathbb{S}_{23}\vert^2 + \vert\mathbb{S}_{13}\vert^2} \equiv s_1 \, ,\qquad 
s^2 >  \dfrac{r^2\vert \mathbb{S}_{23}\vert^2 + \vert \mathbb{S}_{13}\vert^2}
{\vert \mathbb{S}_{23}\vert^2 + \vert\mathbb{S}_{13}\vert^2} \equiv s_2^2\, . 
\label{A1}
\end{equation}
Using (\ref{cond1}) it is not difficult to show that $s_2 > s_1 > r$. Therefore, 
\begin{eqnarray}
J_3^Q <0 \, ,   & \quad   {\rm for} \quad r<s<s_1\, , 
\label{A2}\\
J_3^Q >0 \, ,   & \quad   {\rm for} \quad s_2 < s<1\, .  
\label{A3}
\end{eqnarray}
For $s\in (s_1,s_2)$ the sign of $J_3^Q$ depends on $\lambda $ as well. In this way one finds 
\begin{equation} 
\eta (\lambda;r,s) = 
\begin{cases}
\frac{A}{B_1} & \quad   {\rm for} \quad r<s<s_1\, , \\ 
\frac{A}{B_2} & \quad   {\rm for} \quad s_2<s<1\, , 
\label{A4}
\end{cases}
\end{equation} 
where 
\begin{equation}
A=\lambda \ln \left (1+\e^{-\lambda} \right )\left[|\S_{12}|^2 (1-r)^2 + |\S_{13}|^2(1-s)^2+|\S_{23}|^2 (s-r)^2\right ]\, , 
\label{A5}
\end{equation}
\begin{eqnarray}
B_1=\lambda \ln \left (1+\e^{-\lambda} \right ) \left[|\S_{12}|^2 (1-r) + |\S_{13}|^2 (1-s)\right ] - \nonumber \\
\li_2 \left (-\e^{-\lambda} \right )\left[|\S_{12}|^2 (1-r^2) + |\S_{13}|^2 (1-s^2)\right ]\, , \qquad  
\label{A6}
\end{eqnarray}
and 
\begin{eqnarray}
B_2 = 
\lambda \ln \left (1+\e^{-\lambda} \right ) \left[|\S_{12}|^2 (1-r) + |\S_{13}|^2 (1-s)^2+|\S_{23}|^2 s(s-r)\right ] - 
\nonumber \\
\li_2 \left (-\e^{-\lambda} \right )\left[|\S_{12}|^2 (1-r^2) + |\S_{23}|^2 (s^2-r^2)\right ]\, . \qquad \qquad \qquad \quad 
\label{A7}
\end{eqnarray}



\end{document}